# Polariton Bose-Einstein condensate at room temperature in a Al(Ga)N nanowire-dielectric microcavity with a spatial potential trap

Ayan Das[1], Pallab Bhattacharya[1,*], Junseok Heo[1], Animesh Banerjee[1] and Wei Guo[2]

[1]*Center for Photonic and Multiscale Nanomaterials, Department of Electrical Engineering and Computer Science, University of Michigan, Ann Arbor, Michigan 48109, USA*
[2]*Department of Electrical and Computer Engineering, University of Michigan, Dearborn, 48128, USA*
[*]Corresponding author: pkb@eecs.umich.edu, Phone: 734-763-6678; Fax: 734-763-9324

**Abstract**

A spatial potential trap is formed in a 6.0 μm Al(Ga)N nanowire by varying the Al composition along its length during epitaxial growth. The polariton emission characteristics of a dielectric microcavity with the single nanowire embedded in-plane has been studied at room temperature. Excitation is provided at the Al(Ga)N end of the nanowire and polariton emission is observed from the lowest bandgap GaN region of the nanowire. Comparison of the results with those measured in an identical microcavity with an uniform GaN nanowire and having an identical exciton-photon detuning suggests evaporative cooling of the polaritons as they are transported across the trap in the Al(Ga)N nanowire. Measurement of the spectral characteristics of the polariton emission, their momentum distribution, first-order spatial coherence and time-resolved measurements of polariton cooling provide strong evidence of the formation of an equilibrium Bose-Einstein condensate, a unique state of matter in solid state systems, in the GaN region of the nanowire, at room temperature. An equilibrium condensate is not formed in the GaN nanowire dielectric microcavity without the spatial potential trap.

**Introduction**

Bosons are fundamental particles in nature having integral spin, with occupation in equilibrium characterized by Bose-Einstein statistics, and exhibiting a phenomenon known as bosonic final state stimulation, or Bose stimulation. The latter produces stimulated emission and gain in a conventional laser. It can also produce an equiphase degenerate condensate in equilibrium, termed a Bose-Einstein condensate (BEC)[1]. A BEC characteristically demonstrates long term temporal coherence and long range spatial coherence. In solid state systems, excitons behave as bosons when the mean distance between excitons is much larger than the exciton Bohr radius. Owing to their light mass, it is possible to observe Bose condensation at standard cryogenic temperatures and the possibility of observing an excitonic BEC has been explored with $Cu_2O$[2-4], $CuCl$[5,6] and with GaAs-AlGaAs coupled quantum wells[7-9]. An even more attractive possibility of observing a BEC in a solid state system is with exciton-polaritons[10-12] which are formed in a microcavity by strong coupling between a cavity photon and an exciton. Polaritons, have the extremely small mass of the cavity photon ($m_{pol} \sim 10^{-8}\, m_{atom}$), giving rise to the possibility of forming a BEC at much higher temperatures, even at 300K. In order to form a degenerate condensate, polaritons with higher energy relax by a combination of polariton-phonon and polariton-polariton scattering and eventually condense by phase coherent bosonic final state stimulation. At the same time, polaritons are lost from the system as photons which have a small lifetime. The polariton gas can be in self-equilibrium, or quasi equilibrium, when the polariton relaxation time and lifetime are comparable and the polariton temperature remains larger than the ambient (lattice) temperature. Coherent light output, generally described as polariton lasing[13-18], can be observed under these conditions. A true BEC in perfect equilibrium would be formed by dynamic condensation only if the polariton relaxation time is much shorter than the polariton (photon) lifetime. In such a polariton BEC, the polariton gas should be in equilibrium with the lattice. A large number of pioneering experiments have been performed to realize a BEC with cavity polaritons[19-21]. These observations have been mostly limited to cryogenic temperatures. However, by engineering the polariton scattering rates and lifetimes by choice of materials and the exciton-cavity detuning and by employing evaporative cooling techniques[21], it may be possible to achieve a state of matter - uniquely in a solid-state system - that can exhibit thermal equilibrium at room temperature. Because of the loss through the photonic component, which also provides the opportunity to probe the system, the exciton-polariton system will exhibit this equilibrium condensate only for a short period of time.

GaN with a bandgap of 3.4 eV has a large exciton binding energy of 26 meV. Hence a large Rabi splitting of 56 meV[18] and a trap depth larger than $k_B T$ have been measured at room temperature for GaN-based quantum well[17] and nanowire microcavities[18]. The Bohr radius of GaN is significantly smaller than that of GaAs. Therefore, at room temperature the polariton-phonon scattering rates are large, and no relaxation bottleneck is observed in this system. Baumberg et al[22] have reported spontaneous symmetry breaking of the polarization of polaritons at room temperature in a GaN device. We have recently reported the observation of ultra-low threshold polariton lasing at room temperature[19] and dynamic condensation at ~ 90K[23] in a single GaN nanowire-microcavity device. GaN nanowires grown on Si are relatively free of extended defects and mismatch strain[24-30]. Additionally, a modified cavity field that concentrates within the nanowire ensures a strong interaction between the light and the excitons[18]. Use of the nitride material system, and nanowires in particular, presents the opportunity of observing a polariton condensate in thermal equilibrium at room temperature. The design of the nanowire embedded in the dielectric microcavity is illustrated in Fig. 1. The nanowire is grown on silicon and consists of graded Al(Ga)N along the length. Thus a spatial potential trap is created. If the nanowire is selectively excited at the $Al_{0.05}Ga_{0.95}N$ end, the exciton-polaritons which are more photon-like will relax to lower energies by scattering as they drift and diffuse to the central GaN region where they are less photon-like. The polariton-phonon scattering rate will gradually increase due to the increasing exciton-photon detuning in the microcavity. At the same time the higher energy polaritons (at higher $k_∥$ values) will be lost as photons, thereby providing evaporative cooling. The coldest polaritons near $k_∥ \sim 0$ will reach the bottom of the potential trap in the GaN region. In this paper we report the results of strong coupling experiments performed with such a nitride-based compositionally graded nanowire microcavity. We demonstrate the formation of a polariton BEC in thermodynamic equilibrium at room temperature.

**Results and Discussion**

The Al(Ga)N nanowire consists of a 2.5 μm $Al_{0.05}Ga_{0.95}N$ region, followed by 2 μm graded Al(Ga)N, 1 μm GaN and 0.5 μm graded Al(Ga)N terminating in $Al_{0.05}Ga_{0.95}N$ (Fig. 1(a)). The spatial potential profile that is formed is therefore not symmetric. The Al composition at the end of the nanowire was estimated to be 5% from the measured exciton luminescence peak[31] of 356 nm (3.483 eV) (Fig. 1(a)) and from independent X-ray diffraction measurements. The average polariton diffusion length for this range of alloy composition is estimated to be 61 μm (see supplementary information). The energy difference between $Al_{0.05}Ga_{0.95}N$ and GaN, or the height of the spatial potential trap, is 85.7 meV over a distance of ~ 4.5 μm. The single nanowire is positioned in the microcavity such that ~ 1μm of the $Al_{0.05}Ga_{0.95}N$ region is exposed from the mesa and optical excitation is provided at this end (Fig. 1(b)). We will henceforth refer to this sample with the spatial trap as the Al(Ga)N nanowire microcavity (sample 1). A sample consisting of a single GaN nanowire- microcavity (sample 2) having the same cavity photon-exciton detuning δ was also fabricated for comparison of experimental results. Micro-photoluminescence spectra from the GaN nanowire exhibits the three free excitons $X_A$, $X_B$ and $X_C$ and the corresponding donor bound (DB) transitions. Of these the $X_A$ and $DBX_A$ are the most dominant in the spectra, and hence coupling of the $X_A$ exciton to the cavity mode is only considered for analyzing measured polariton dispersion characteristics with the coupled oscillator model[18].

The polariton dispersion characteristics of the Al(Ga)N nanowire microcavity was determined from measured angle resolved PL (0°-30°) data recorded for low excitation densities, which show strong polariton luminescence corresponding to the GaN (center) region of the nanowire. The results were analyzed with a coupled oscillator model taking into account the coupling of the $X_A$ exciton and the cavity mode and values of Rabi splitting $\Omega$ = 48 meV and cavity-to-exciton detuning δ = +2 meV are derived. The same value of δ is derived for sample 2 with the single GaN nanowire.

The integrated intensity of polariton luminescence recorded at room temperature in the normal direction ($k_\parallel \sim 0$) as a function of pump power exhibits a non-linear behavior, accompanied by a sharp decrease of the emission linewidth. Figures 2 (a) -(c) illustrate these characteristics for the Al(Ga)N nanowire microcavity (sample 1). Similar data recorded for the GaN nanowire sample (sample 2) is illustrated in Fig. 2(d). The incident excitation energies at the threshold of the non-linearity are 102 nJ/cm$^2$ and 125 nJ/cm$^2$ for sample 1 and 2, respectively, and the minimum emission linewidths are 0.43 meV and 1.1 meV respectively. It may be noted that the reduction in threshold power and emission linewidth agrees with the observation made by Balili et al[32] where similar trends were recorded for an increase in the locally applied stress (trap depth) for the GaAs/AlGaAs QW microcavity. The emission spectra above threshold (Fig. 2(a)) exhibits multiple peaks with narrow linewidth corresponding to different discrete LP transverse modes localized in the nanowire. Figure 2(c) shows the variation of the LP and upper polariton (UP) peak energies. The extremely small shift in energy of the LP mode (~1.5 meV, compared to $\Omega$ = 48 meV) indicates that the system remains in the strong coupling regime over the entire range of excitation energy. In addition, as seen in Fig. 2(b) for sample 1, the linewidth of 0.43 meV that persists for a range of excitation is at the resolution limit of the spectrometer. Therefore, the emission linewidth could be smaller than 0.43 meV. The coherence times corresponding to the minimum measured emission linewidths are 9.6 ps (sample 1) and 3.75 ps (sample 2). The upper bounds of the LP density at threshold are $2.2 \times 10^{16}$ cm$^{-3}$ (sample 1) and $2.7 \times 10^{16}$ cm$^{-3}$ (sample 2), using the relation $N_{3D} < E_{th1}/(E_{pump}D)$. These polariton densities are 3 orders of magnitude smaller than the exciton Mott density[33] of $3 \times 10^{19}$ cm$^{-3}$. Furthermore, in a single GaN nanowire microcavity device identical to sample 2, we have measured a value of $E_{th}$ = 92 nJ/cm$^2$ at 300K. We also measured conventional photon lasing in the same device with an energy threshold 2700 times larger in value[18]. While these characteristics indicate coherent emission from a degenerate and coherent LP condensate, or polariton lasing, they do not confirm that the condensate is in equilibrium with the lattice. It is evident, however, that the incorporation of the spatial potential trap produces noticeable differences in the characteristics of polariton emission.

In order to gain a better understanding of the polariton dynamics and the occupation of the LPs in $k_\parallel$ space, we have performed angle-resolved PL measurements as a function of excitation density for samples 1 and 2. It may be noted that pulsed excitation has been used and hence the polariton density and temperature may change with delay after the excitation pulse. However, for excitation powers $P \leq 1.7P_{th}$, the polariton relaxation time remains larger than or comparable to the polariton emission time constant and the distribution of LP density in $k_\parallel$-space can be considered invariant[14,23]. Under these conditions a time-integrated emission intensity obtained from the pulsed excitation measurement is a good approximation for the LP density. At higher excitation powers, this approximation is not valid due to the higher polariton scattering rates. Below the threshold for non-linear emission (Fig. 3) the PL spectra remains broad and of comparable intensity for all angles. Above threshold the emission is characterized by a spectrally narrow and strong peak at $k_\parallel \sim 0$. The relative occupancy, or the LP number density per $k_\parallel$-state, at 300K is plotted against $E-E(k_\parallel \sim 0)$ for different excitation powers in Figs. 4(a) and (b) for the Al(Ga)N nanowire (sample 1) and GaN nanowire (sample 2) microcavities, respectively. Far below the non-linear threshold the polariton scattering rate is inadequate to relax the polaritons and the LP distribution remains non-thermal. Near and above threshold, the relaxation dynamics is fast enough and a dynamic condensation leading to a higher occupancy at or near ground state ($k_\parallel \sim 0$) is observed. At the higher excitation powers a bimodal distribution of the occupancy accompanied by a sharp increase in LP occupancy near $k_\parallel \sim 0$ is observed for sample 1 having the spatial potential trap. In the range $(0.9-1.0)P_{th}$, the occupation distribution in $k_\parallel$-space can be described by the Maxwell-Boltzmann (MB) distribution. At and above threshold, a Bose-Einstein (BE) distribution: $N_{BE}(k) = 1/[\exp(E/k_B T_{LP})(1+ N_0^{-1}) - 1]$ can be used to analyze the polariton occupation data. For example for $P = 1.0P_{th}$, $1.1P_{th}$ and $1.3P_{th}$ in sample 1, values of $T_{LP} = 298.1K$, $300.7K$ and $303.4K$ are obtained from a fit to the data. At higher excitation levels ($P=1.5P_{th}$), thermal equilibrium with the phonon bath is lost, as the hot polaritons generated by the excitation cannot be adequately cooled, and $T_{LP}$ increases[23]. The significant bimodal polariton distribution of the occupancy $N(Ek_\parallel)$ at high excitation levels in sample 1 is predicted from steady-state quasi-equilibrium theory[34,35] and is a strong indication of the formation of a Bose condensate. In contrast, a condensate in thermal equilibrium ($T_{LP} \cong T_{lattice}$) and a distinct bimodal distribution are not observed for sample 2 without the spatial potential trap, although the detuning is the same ($\delta = +2$ meV). A value of $T_{LP} = 357$K is derived for $P = 0.9P_{th}$ for sample 2. For an ideal BEC, the chemical potential $\mu$ given by $-\mu/k_BT = \ln(1+N_0-1)$ tends to zero. The fit to the occupation data of sample 1 with a BE distribution for $P = 1.1P_{th}$ yields $\mu = -4.7$ meV. Furthermore, as the false color plots of Fig. 3 indicate, at this excitation, the condensate at $k_\parallel \sim 0$ is described by extremely small values of $\Delta k \sim 1 \times 10^4$ cm$^{-1}$ and $\Delta E \sim 0.5$ meV.

Time-resolved photoluminescence (TRPL) measurements have been made with the 267 nm pulsed excitation and a streak camera with an overall temporal resolution of 5 ps. The transient data recorded at room temperature for sample 1 are shown in Fig. 5(a). The rising part of the transient, which principally corresponds to the filling of the exciton reservoir in the Al$_{0.05}$Ga$_{0.95}$N region of the nanowire with excitation, is limited by system resolution. The decay time decreases with increasing excitation, reflecting enhanced polariton relaxation into the $k_\parallel \sim 0$ states in the GaN (center) region of the nanowire. The transport and scattering of the polaritons from the exciton reservoir in the Al$_{0.05}$Ga$_{0.95}$N region of the nanowire, across the spatial potential trap and into the bottom of the LPB ($k_\parallel \sim 0$) in the GaN region, was explained earlier and illustrated in Fig. 1(a). This complex dynamic process can be analyzed by detailed Monte Carlo simulation to determine the polariton scattering times at each point in the trap, which is beyond the scope of the present work. Instead, we describe the evaporative cooling due to emission of hot polaritons escaping as photons, as the polaritons are transported along the trap, and the population redistribution in momentum space due to polariton-phonon and polariton-polariton scattering, at each point in the trap, by a single time constant $\tau_{therm}$ in the coupled rate equations:

$$\frac{dn_R}{dt} = P(t) - \frac{n_R}{\tau_R} - \frac{n_R}{\tau_{therm}} \quad (1)$$

$$\frac{dn_0}{dt} = -\frac{n_0}{\tau_0} + \frac{n_R}{\tau_{therm}} \quad (2)$$

Here P(t) represents the excitation as a Gaussian pulse centered at t = 0, $n_R$ and $n_0$ represent the populations in the $Al_{0.05}Ga_{0.95}N$ reservoir and the condensate at $k_\parallel \sim 0$ in the GaN region, respectively, and $\tau_0$ is the polariton lifetime in this condensate. The exciton lifetime in the reservoir, $\tau_R$, is usually much larger (~ 1 ns) than $\tau_0$ and therefore the corresponding term in the rate equation is neglected. The values of $\tau_0$ (= 0.55 ps) is fixed at $k_\parallel \sim 0$, given by $\tau_0 = \tau_{ph}/|Ck_{\parallel=0}|^2$ where $\tau_{ph} \sim 0.26$ ps determined from the cavity Q ~ 600 and $|Ck_{\parallel=0}|^2$ is the photon fraction in the LPB at $k_\parallel \sim 0$ (Fig. 1(a)). Plotted in Fig. 5(b) are the values of $\tau_{therm}/\tau_0$ against the relative excitation power. It is evident that the value of $\tau_{therm}$ decreases rapidly at $P/P_{th} \sim 1$ and reaches $0.09\tau_0$ at the highest excitation power. Similar TRPL measurements were also made on the GaN nanowire microcavity (sample 2) and the transient data were again analyzed by the coupled rate equations (1) and (2). In this case $\tau_{therm}$ represents the overall scattering (relaxation) time from the reservoir to the condensate at $k_\parallel \sim 0$, neglecting the non-linear time constants and polariton densities characterizing the intermediate polariton scattering processes. The variation of $\tau_{therm}/\tau_0$ for sample 2 is also shown in Fig. 5(b) and it is evident that $\tau_{therm}$ tends to saturate at a value comparable to that of $\tau_0$ ( $0.6\tau_0$). The steep decrease of $\tau_{therm}/\tau_0$ observed for sample 1 has not been reported for any other microcavity polariton emitter. From the measured linewidths of polariton emission in sample 1, the coherence time $\tau_{coherence}$ increases from 0.34 ps below threshold to a value of ~ 10 ps just above threshold. This is significantly larger than polariton lifetime ($\tau_0$) of 0.55 ps at $k_\parallel \sim 0$. In turn, $\tau_0$ is larger than $\tau_{relax}$ by an order of magnitude above threshold. We therefore believe that the incorporation of graded Al(Ga)N in the nanowire, forming a potential trap and enabling evaporative cooling, assists the dynamic condensation process and singularly creates an equilibrium phase transition in the Bose condensate at 300K. In contrast, in the GaN nanowire sample, polariton-phonon scattering is not sufficient enough to cool the polaritons to form a BEC at 300K.

Additional characteristics of a BEC include long-range coherence and spontaneous symmetry breaking of the polarization of the polaritons. We have measured the first-order correlation of the polariton luminescence with a Michelson interferometer. The cw excitation source is a 325 nm HeCd laser and the incident power is varied from 0.3 mW to 1.5 mW. The experimental arrangement, schematically shown in the supplementary material, creates a double image of the condensate with a spatial resolution of 0.3 µm. Figure 6 shows the measured contrast $(I_{max} - I_{min})/(I_{max} + I_{min})$ of the fringes as a function of displacement between the images in the two arms. The measurement therefore provides direct evidence of coherence in the first-order correlation of the polariton emission and an estimate of the spatial coherence length of the condensate along the length (or c-axis) of the nanowire. Below threshold, the contrast remains within the noise margin for any overlap between the images, indicating that there is negligible coherence across the condensate. Above threshold, a maximum contrast of 21% is observed at P = 1.1 mW for complete overlap between the two images and the contrast decreases as the images are moved apart. The linewidth (FWHM) of the contrast profile is ~ 1.2 µm, which is a measure of the size of the condensate formed in the trap above threshold and it is approximately the same as the length of the GaN region in the nanowire.

The polarization of the polariton emission was also measured as a function of incident power and the output was linearly polarized below and above threshold, unlike the case of bulk/quantum well planar microcavities, where spontaneous symmetry breaking is responsible for a linearly polarized emission above threshold[22]. The measured polarization is shown in the supplementary information and the results can be understood in the context of the geometry of our sample shown in Fig. 1(b) and in the supplementary information. The nanowire is embedded in-plane in the microcavity. The $X_A$ exciton-polariton emission is collected in the direction vertical to the c-axis of the nanowire. The emission must therefore be linearly polarized in a direction perpendicular to the c-axis and the direction of emission, independent of the excitation intensity. Polariton superfluidity and associated quantized vortices observed in exciton-polariton condensates[36,37] and dilute atomic gas BECs[38,39] could not be investigated because of the intrinsically weak emission from a 1 µm length of GaN in the nanowire microcavity which limits any imaging of the interference pattern and subsequent extraction of phase information.

In conclusion, a spatial polariton trap is created along the length of a Al(Ga)N nanowire and strong coupling effects have been investigated in the polariton emission of a planar nanowire-dielectric microcavity sample. It is evident that the trap assists in evaporative cooling of the optically excited exciton-polaritons as they are transported along the nanowire, to form an equilibrium BEC in the lowest bandgap GaN region of the nanowire at 300K.

**Materials and Methods**

The experimental sample consists of a single graded bandgap Al(Ga)N nanowire of length equal to 6 µm and diameter ~ 50 nm embedded in a $Si_3N_4$ λ-cavity surrounded by $SiO_2$/ $Si_3N_4$ distributed Bragg reflectors (DBRs) on top and bottom. The nanowire sample is grown by molecular beam epitaxy on (001) silicon substrate and has the wurtzite crystalline form with the c-axis parallel to the growth direction. The nanowires are dispersed on half of the $Si_3N_4$ cavity and single nanowires are selected by e-beam lithography. Square shaped mesa microcavities of 10 μm side are formed by photolithography and etching on a Si substrate.

Time-integrated photoluminescence (PL) measurements were performed on the microcavity samples by non-resonant excitation with the linearly polarized output of a frequency tripled ($\lambda$ = 267 nm, $f_{rep}$ = 80 MHz and pulse width of 150 fs) Ti:sapphire laser. A doublet lens was used to focus the pump beam at an incident angle of 20º on to the exposed $Al_{0.05}Ga_{0.95}N$ end of the nanowire. The excitons created by optical absorption enter the microcavity and spontaneously form exciton-polaritons which then drift and diffuse to the center GaN region of the nanowire. However, this arrangement does not completely prevent light backscattered by the lower DBR from exciting the entire nanowire. A finite difference time domain (FDTD) calculation, taking into account the excitation and sample geometry, indicates that the intensity of excitation by the scattered light in the GaN region is approximately 1% of the incident excitation intensity (see supplementary information). Hence the number of exciton-polaritons directly generated in the GaN region of the nanowire is negligible. The luminescence was collected from the center GaN region of the nanowire by a 1.5 μm core fiber patch cord and transmitted to a spectrometer (with a spectral resolution of ~ 0.05 nm). The collection optics is located on extended rails of a goniometer centered at the sample and has an angular resolution of 1°. In the case of samples with a GaN nanowire, the entire mesa-shaped microcavity is uniformly illuminated by the same laser.

**Acknowledgements**

The work has been supported by the National Science Foundation MRSEC Program under Grant DMR-1120923. The authors gratefully acknowledge the help provided by H.Deng and L.Zhang.

**Figure Legends:**

*Figure 1* (a) Schematic of the Al(Ga)N nanowire showing the variation of the exciton energy as a function of position. The exciton energies at the AlGaN end and the GaN region are estimated from the measured photoluminescence peaks as shown. Photon and exciton fraction of the lower polariton branch at $k_\parallel \sim 0$ are also shown as a function of position in the trap; (b) schematic representation of the dielectric microcavity with a single Al(Ga)N nanowire of diameter 50 nm and length 6 µm buried in the center of a λ-sized cavity. The microcavity is etched into mesas of size ~ 10 µm with 1 µm of the $Al_{0.05}Ga_{0.95}N$ end of the nanowire exposed; (c) high-resolution transmission electron microscope image of a GaN nanowire reveals that the nanowires are relatively free of extended defects and stacking faults. Inset shows the selected area diffraction pattern confirming that the nanowires have a wurtzite structure and grow along the c-axis.

*Figure 2* (a) Emission spectra measured as a function of incident energy density for the Al(Ga)N nanowire device (sample 1); (b) variation of integrated emission intensity and emission linewidth as a function of injection energy density measured for sample 1; (c) variation of upper and lower polariton energies in sample 1 with injection energy density, as deduced from excitation dependent photoluminescence; (d) variation of integrated emission intensity and emission linewidth versus injection energy density measured for the single GaN nanowire device (sample 2).

*Figure 3* Momentum distribution of polaritons at different excitation levels obtained from angle-resolved measurements and displayed as false color plots. The horizontal axis displays the the in-plane momentum and the vertical axis displays the emission energy in a false-color scale. The false color scale is linear with red and blue representing high and low values, respectively. The exciton and cavity photon energies are also indicated.

*Figure 4* Polariton occupancy plotted as a function of energy in a semi-logarithmic scale for various normalized excitation powers for (a) the Al(Ga)N nanowire device (sample 1) and (b) the uniform GaN nanowire device (sample 2). For each excitation power, the zero of the energy scale corresponds to the energy of the $k_{||} \sim 0$ state.

*Figure 5* (a) Transient polariton emission measured as a function of the incident energy density in the Al(Ga)N nanowire device (sample 1); (b) normalized plot of the thermalization time as a function of the incident power obtained from a solution of a two-level rate model (see text).

*Figure 6* Interference fringe contrast measured as a function of the displacement between a double image of the polariton emission for two incident cw pump powers in the Al(Ga)N nanowire device (sample 1). Inset shows measured intensity variation at zero displacement as a function of relative phase between double images for $P_{inc} = 1.2$ mW.

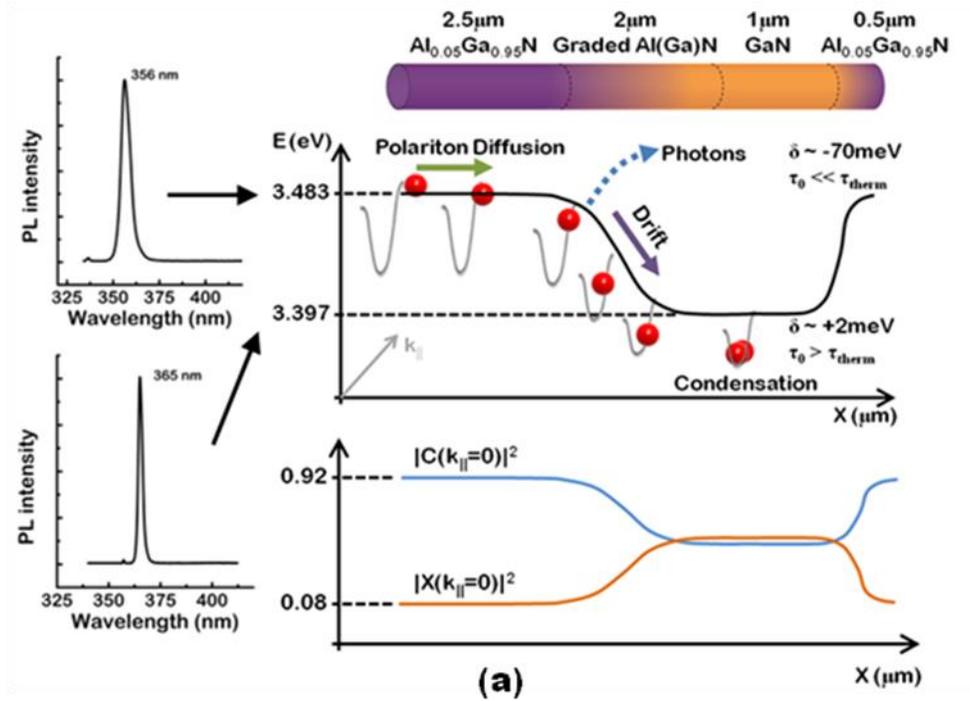
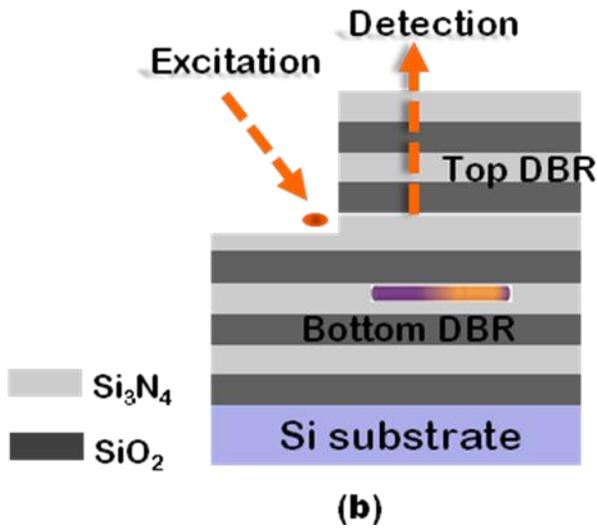
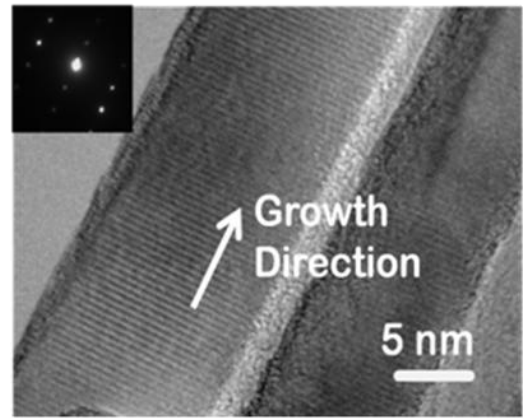

Figure 1 of 6

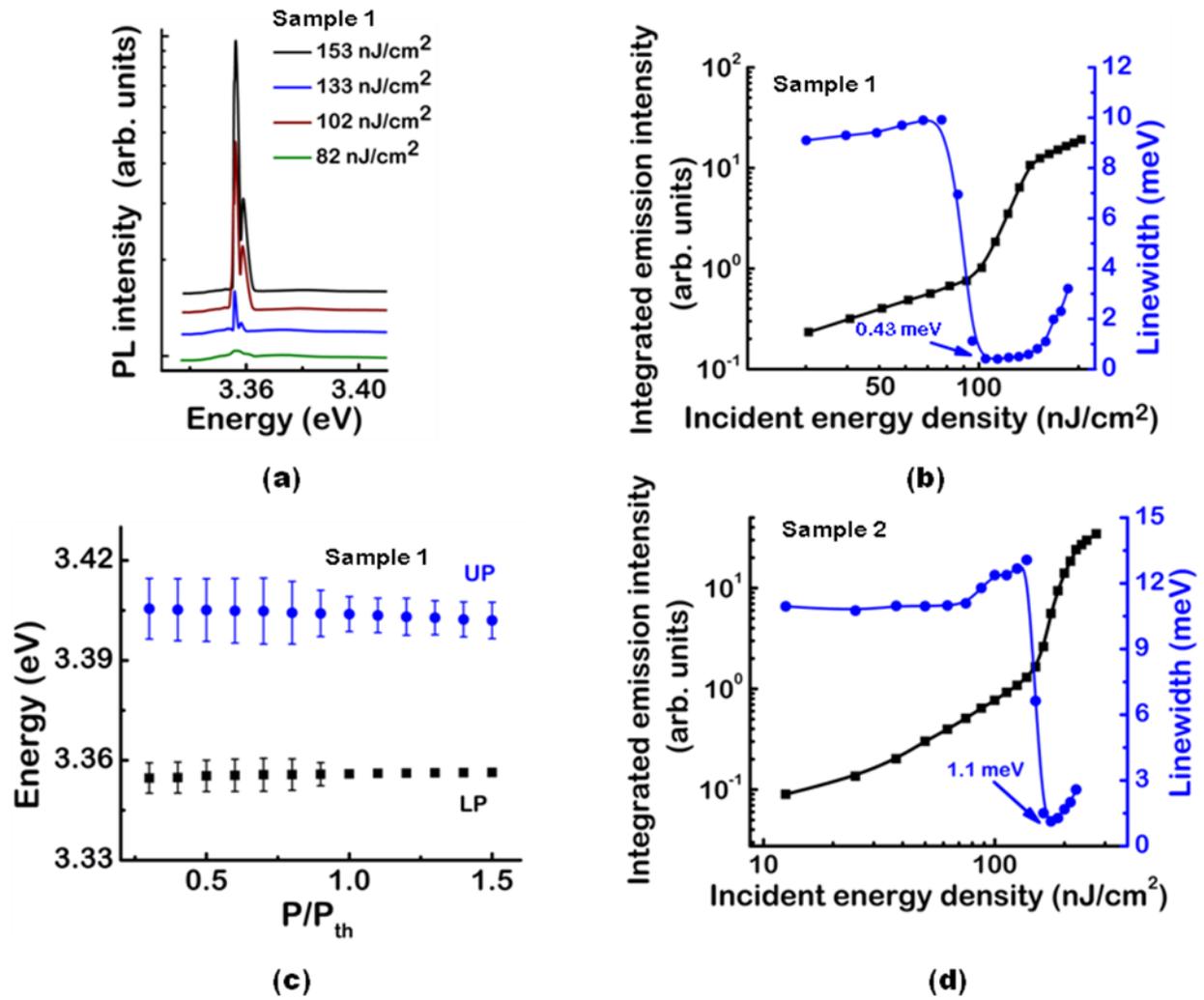

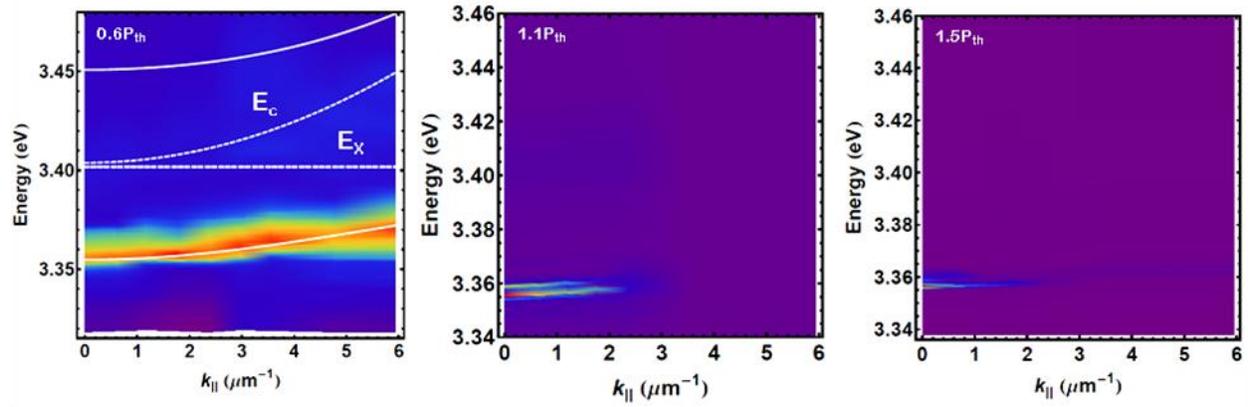

Figure 3 of 6

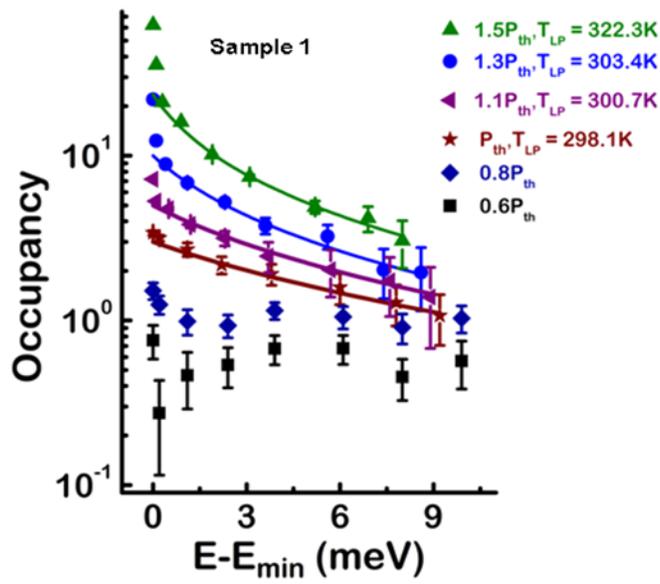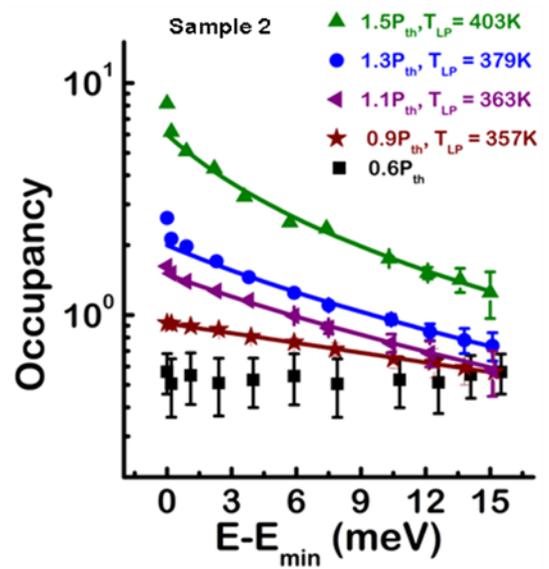

Figure 4 of 6

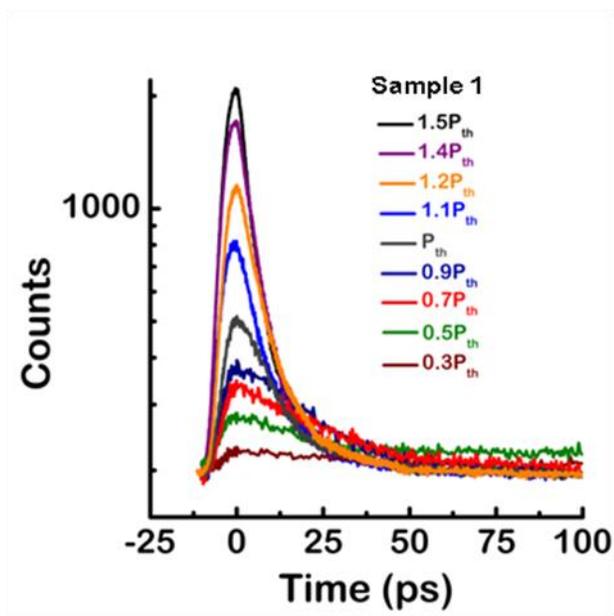 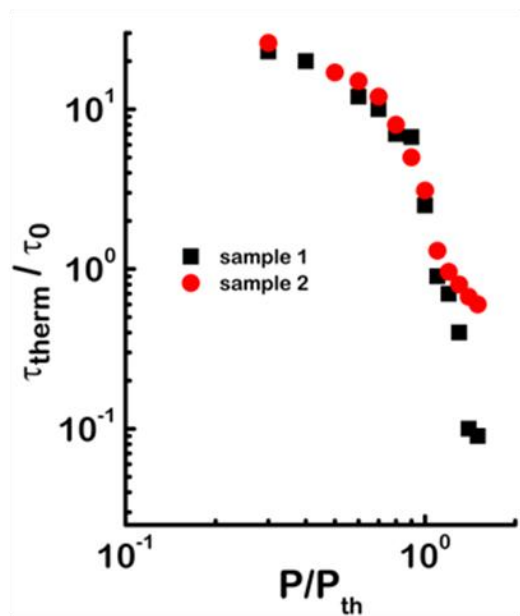

(a) (b)

Figure 5 of 6

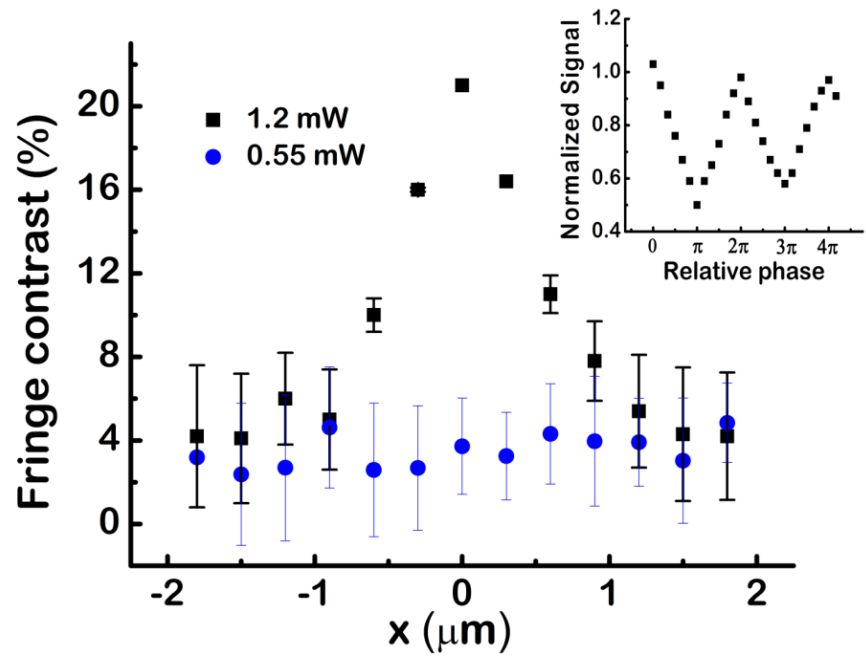

**Supplementary Information**

*1. FDTD simulation to estimate the direct excitation in the buried GaN region:*

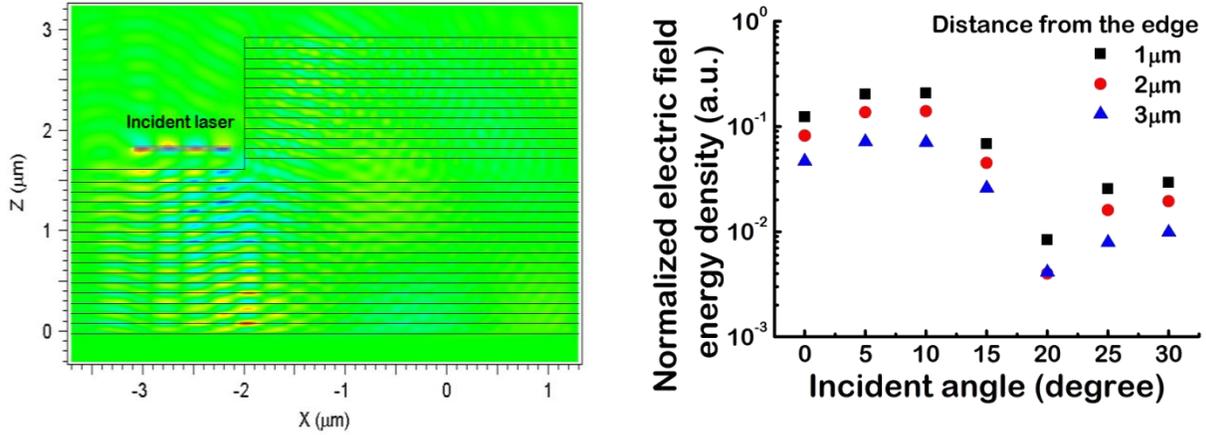

The edge of the mesa was excited by the pump laser at an oblique angle. However, in this geometry, one cannot rule out the contribution from the scattered pump laser guided into the mesa that might excite the entire nanowire. Hence, a finite difference time domain (FDTD) calculation was carried out to investigate the extent of such direct excitation. In the simulation, the source with a wavelength of 267 nm was placed near the edge of the mesa and absorption in the distributed Bragg reflectors was taken into account. The relative energy density inside the mesa with respect to that just below the source was recorded at various incidence angles. It is observed that the direct excitation of the nanowire inside the mesa is at a minimum when the pump laser excitation is at 20° with respect to the normal. The calculated light intensity inside the mesa is only ~1% of the direct excitation.

## 2. Calculation of polariton diffusion length:

### A. Evaluation of average polariton mass -

$\frac{1}{m} = \frac{|X|^2}{m_{exc}} + \frac{|C|^2}{m_{cav}}$, since $m_{cav} \ll m_{exc}$ $m \cong \frac{m_{cav}}{|C|^2}$, where $\frac{1}{|C|^2} = \frac{1}{n}\sum_{i=1}^{n}\frac{1}{|C_i|^2}$ is the inverse of the photon fraction averaged along the trap and the cavity mass $m_{cav} = \frac{E_{cav}(k_{//}=0)}{c^2/n_c}$, where $E_{cav}(k_{\|} = 0)$ is the cavity energy at zero in-plane vector and $n_c$ is the refractive index of the cavity material. Upon calculation $m = 2.53 \times 10^{-5} m_0$ where $m_0$ is the free electron mass.

### B. Evaluation of average polariton lifetime in the ground state -

$\tau_0 \cong \frac{\tau_{cav}}{|C|^2} = 0.44\,ps$, where $\tau_{cav}$ is the cavity photon lifetime estimated from the cavity Q.

### C. Diffusion length -

In two dimensions, the diffusion length is given by $l_D = \sqrt{D\tau_0}$, where the diffusion constant D is given by

$D = \frac{2k_B T}{m}\tau_{sc}$, where $\tau_{sc}$ is the scattering time of polaritons with phonons and is equal to 20 ps in GaN

at room temperature. The estimated diffusion length is equal to 61 μm.

*3. First-order coherence measurements:*

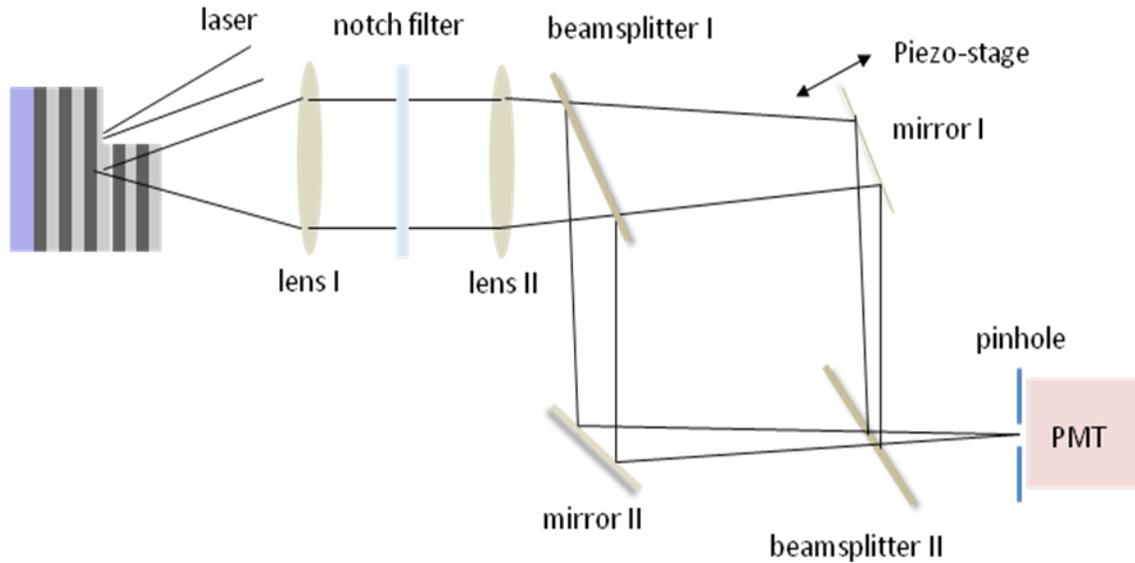

We measured the first order spatial coherence using a Michelson interferometer arrangement as shown above. The beam splitter creates identical images of the condensate in the two arms of the interferometer and they are overlapped spatially by the translational movement of a mirror (mirror I) mounted diagonally on a piezo-stage. The arrangement creates a double image of the condensate with a spatial resolution of 0.3 μm. A pin-hole placed at the overlap image plane serves as a spatial filter across the dark and bright fringes, the intensities of which, and subsequently the contrast, are measured by a photomultiplier tube. The interference pattern could not be imaged with a CCD camera because of the low intensity of the polariton emission. The notch filter with a cut-off at 350 nm ensures that the interference contrast is not coming from laser light scattered off the sample in the normal direction.

*4. Polarization of the emission:*

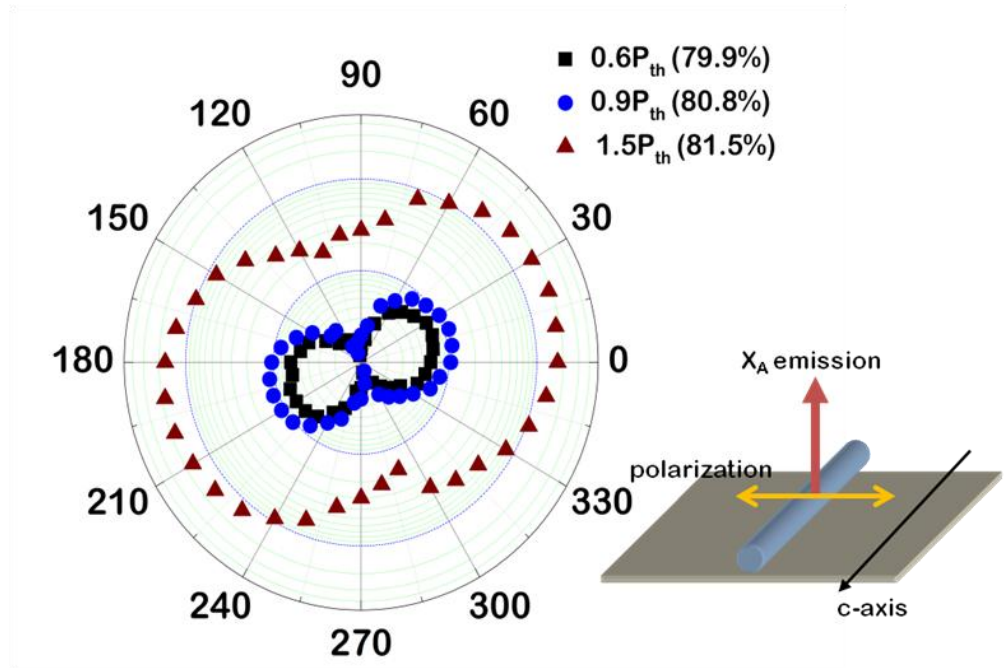

The figure above plots the intensities of the polariton emission measured as a function of the angle of the linear polarizer at 3 different excitation powers. The emission is linearly polarized both below and above threshold. In this geometry, with the nanowire lying in-plane and the $X_A$-polariton emission collected in a direction vertical to the c-axis of the nanowire, the output will always be linearly polarized. It may be noted that the polarization of $X_A$ emission should be perpendicular to the direction of light output and the c-axis simultaneously.